\documentclass [11pt] {article}
\usepackage[dvips]{graphicx,color}
\usepackage{amsmath}
\usepackage[hyperindex,pdfmark,letterpaper]{hyperref}
\newcommand{\bfy}{{\bf y}}
\newcommand{\bfone}{{\bf 1}}

\newcommand{\calk}{{\cal K}}

\definecolor{Red}{rgb}{1,0,0}
\definecolor{Blue}{rgb}{0,0,1}

\usepackage{verbatim}

\begin{document}
\title{
   \bf On the dependence of the Navier-Stokes equations on the distribution of molecular velocities}
\author{
  Peter J. Love, Bruce M. Boghosian\\  
{\footnotesize Department of Mathematics, Tufts University, Medford, Massachusetts 02155 USA}\\
}
\maketitle
\begin{abstract}
In this work we introduce a completely general Chapman Enskog procedure in which we divide the local distribution into an isotropic distribution with anisotropic corrections. We obtain a recursion relation on all integrals of the distribution function required in the derivation of the moment equations. We obtain the hydrodynamic equations in terms only of the first few moments of the isotropic part of an arbitrary local distribution function. 

The incompressible limit of the equations is completely independent  of the form of the isotropic part of the distribution, whereas the energy equation in the compressible case contains an additional contribution to the heat flux. This additional term was also found by Boghosian and by Potiguar and Costa~\cite{bib:bogtsallis,bib:tsallistran} in the derivation of the Navier Stokes equations for Tsallis thermostatistics, and is the only additional term allowed by the Curie principle. 
\end{abstract}
\section{Introduction}
Boltzmann's two crowning achievements were his definition of the microcanonical entropy and Maxwell Boltzmann distribution, and his kinetic equation and mechanical basis for irreversibility of macroscopic non-equilibrium systems. Boltzmann's famous H-theorem gave the first connection of the second law of thermodynamics with underlying reversible mechanics. In this paper we shall revisit both of these contributions. Firstly we shall review the resolution of the apparent conflict between reversible microdynamics and the second law. We then turn to the main subject of this paper, which is the consideration of solutions to the Boltzmann equation when the local velocity distribution is {\em not} a Maxwell-Boltzmann distribution, but an arbitrary isotropic function.

Specification of a set of macroscopic variables obeying an autonomous evolution in agreement with experiment is the aim of any non-equilibrium description of a system. The example we consider here is single-phase compressible fluid mechanics, where the average velocity, density and energy of a vast ($\sim 10^{23}$) number of atoms or molecules moving in $D$ dimensions obey a closed set of equations for $D+2$ macroscopic variables. 

Boltzmann provided us with the first concrete steps towards a general method for finding such closed descriptions. The kinetic equation which bears his name is an approximation to the first equation in the BBGKY hierarchy. In order to obtain a closed equation for the single particle distribution function, Boltzmann made his famous {\it stosszahlansatz}, or {\it molecular chaos} approximation, replacing the two-particle distribution function by a product of one-particle distributions. We may regard the  {\it stosszahlansatz} not as a restriction on the behaviour of microscopic dynamics, with which it is in direct contradiction, but as a restriction on the types of macroscopic variables for which we are seeking autonomous descriptions. That is, we seek autonomous  descriptions for macroscopic variables depending on the single particle distribution function. 

Having obtained the Boltzmann equation, the Chapman Enskog expansion gives us a method for finding the moment equations~\cite{bib:chapmancowl,bib:leb3}. The moment equations, as we shall see, are obtained as a hierarchy (in this case referred to as the super-Burnett hierarchy). The lowest order equations presume that an exact local equilibrium exists everywhere, and the resulting equations are the inviscid Euler equations. At the next level one introduces small deviations from equilibrium, and the resulting moment equations are the Navier Stokes equations. At each successive level in this hierarchy one expresses higher order corrections to the local equilibrium distribution in terms of gradients of that local equilibrium. 

As emphasized by Chapman and Cowling~\cite{bib:chapmancowl}, the division of the distribution into equilibrium and non-equilibrium parts is non-unique. It is therefore unclear what range of validity a given moment equation in the super-Burnett hierarchy has. There are known situations of considerable practical interest in which the Navier Stokes equations must be replaced by a super-Burnett scheme~\cite{bib:agarwal}. In general it is necessary to validate any given scheme against some more fundamental approach such as direct simulation Monte Carlo~\cite{bib:levermore}. Furthermore the entire edifice crumbles if we do not know what equilibrium distribution the fluid will tend to. This has been a particular stumbling block when one wishes to extend the Chapman Enskog treatment to interacting fluids where one does not in general know the equilibrium distribution.

However, even in the case of non-interacting fluids an interesting case presents itself. Tsallis and others have proposed various generalizations of Boltzmann-Gibbs thermostatistics, and experimental studies have verified that such statistics arise in nature~\cite{bib:tsallisrev}. Recent work by Beck and Cohen has placed all such generalized statistics schemes on a common footing by utilising the physically appealing notion of fluctuating {\em intensive} parameters~\cite{bib:superstatistics}. 

From an equilibrium point of view, such thermostatistics are somewhat controversial. It is unclear under what circumstances a system's equilibrium will deviate from the Boltzmann-Gibbs form. Although there are many systems known to so deviate, without a theoretical connection between a systems microscopic properties and its deviations from Boltzmann-Gibbs statistics, such distributions remain unsatisfactorily phenomenological. 
Away from equilibrium, the picture is quite different. Boltzmann's treatment of irreversible processes does not specify a single functional form for the entropy out of equilibrium. His $H$-function is equal to the Gibbs entropy only for an ideal gas in equilibrium. In any other circumstances these quantities are not equal~\cite{bib:gull}. In general, the entropy away from equilibrium is only defined with respect to a set of macroscopic variables undergoing an autonomous evolution. Consideration of the widest possible range of distributions and entropies seems only prudent in this context.

We have emphasized that it is the existence of autonomous evolution equations for macroscopic parameters which defines the subject of non-equilibrium thermodynamics. Single phase fluid mechanics is the canonical example of such an autonomous description, and therefore how this description depends on the details of the local distribution is of some interest. In particular, under what circumstances does the hydrodynamic description cease to be autonomous? Conversely, determining the dependence of the hydrodynamic equations on the details of the distribution may provide experimentalists with the capability to determine the thermostatistics of a system by purely hydrodynamic experiments.

The Boltzmann equation, H-theorem and Chapman-Enskog expansion have previously been studied in the context of Tsallis thermostatistics, and a set of Navier-Stokes equations have been derived for the Tsallis generalized statistics~\cite{bib:bogtsallis,bib:tsallisH,bib:tsallistran}. Boghosian found that the incompressible limit of the Navier-Stokes equations is independent of the parameter $q$ characterizing the Tsallis distribution, whereas the energy equation in the compressible case contains an additional $q$-dependent contribution to the heat flux~\cite{bib:bogtsallis}. This result was confirmed in~\cite{bib:tsallistran}, and the additional term is the only further term allowed by the Curie principle~\cite{bib:degrootmazur}.

We begin in Section~\ref{sec:CE} by constructing the well known Chapman-Enskog expansion for the Boltzmann equation with a BGK collision operator. We then write the first and second order solutions in terms of moments of the distribution function. In Section~\ref{sec:mom} we introduce the ``flux-field'' expansion defined in~\cite{bib:micrork} and the recursion relation proved in~\cite{bib:lovepart}, and use this recursion to evaluate the moments, thereby obtaining the hydrodynamic equations for an arbitrary isotropic local equilibrium, in terms of the low order moments of that distribution. We close the paper with some further discussion and directions for future work. Throughout the following we choose units such that the sound speed $c=1$ and the mass of the molecules composing the fluid $m=1$. In particular this means that the hydrodynamic velocity ${\bf u}$ is given in units of the Mach number.

\section{Chapman - Enskog Expansion}\label{sec:CE}
We begin with the BGK equation in $D$ dimensions:
\begin{equation}\label{eq:bgk}
\frac{\partial \phi}{\partial t} +{\bf v}\cdot \nabla \phi = -\frac{1}{\tau}\bigl(\phi-\phi^{(0)}\bigr).
\end{equation}
Where $\phi$ is the single particle distribution function, $\phi^{(0)}$ is an arbitrary isotropic function, ${\bf v}$ is the molecular velocity, $\tau$ is a parameter (the collision time) and $t$ is time. We define the operator
\begin{equation}
{\cal D} = \frac{\partial}{\partial t} +{\bf v}\cdot \nabla 
\end{equation}
in terms of which we may write the formal solution of~(\ref{eq:bgk}):
\begin{equation}\label{eq:bgksol}
\phi = (1+{\cal D})^{-1}\phi^{(0)}
\end{equation}
We now introduce a hierarchy of separated length and time scales, such that we write the operator ${\cal D}$:
\begin{equation}
{\cal D}=\sum_{n=1}^{\cal 1}\epsilon^n{\cal D}_n=\sum_{n=1}^{\cal 1}\epsilon^n\left(\frac{\partial}{\partial t_n} +{\bf v}\cdot \nabla\right)
\end{equation}
Usually in the Chapman Enskog expansion one considers that relaxation to a local equilibrium happens on timescale $t_1$, and slower kinetic and hydrodynamic effects occur on progressively longer timescales. We wish to consider a slightly more general situation where the distribution relaxes to a locally isotropic form on timescale $t_1$. Subsituting the above in our formal solution~(\ref{eq:bgksol}) we obtain:
\begin{equation}
\phi = \left[1+\sum_{p=1}^{\cal 1}(-1)^p\tau^p\left(\sum_{n=1}^{\cal 1}\epsilon^n{\cal D}_n\right)^p\right]\phi^{(0)} 
\end{equation}
Expanding our distribution about the isotropic part:
\begin{equation}
\phi = \phi^{(0)} + \epsilon \phi^{1} + \dots
\end{equation}
yields a hierarchy of equations for the successive deviations from equilibrium $\phi^n$:
\begin{equation}
\begin{split}
\phi^{(1)}&=-\tau{\cal D}_1\phi^{(0)}\\
\phi^{(2)}&=-\tau({\cal D}_2-\tau{\cal D}_1^2)\phi^{(0)}\\
\phi^{(3)}&=-\tau({\cal D}_3-2\tau{\cal D}_1{\cal D}_2 + \tau^2{\cal D}_1^3)\phi^{(0)}\\
&\dots
\end{split}
\end{equation}
If we additionally specify that the deviations from equilibrium are not to affect the equilibrium values of the moments of the distribution we may obtain a hierarchy of conservation laws for those moments. At order $\epsilon$ we obtain the continuity and Eulers equations for the fluid, at order $\epsilon^2$ we obtain the Navier-Stokes equations for viscous compressible flow and at subsequent orders we obtain the Burnett equations and super-Burnett equations. In the present paper we consider only the first two orders in this hierarchy, but note in passing that the techniques employed here are equally applicable to obtaining such higher order hydrodynamic schemes.
\subsection{First order solution}
Multiplying the first order equation by the mass, momentum and kinetic energy of the fluid particles and integrating over velocity gives:
\begin{equation}
\begin{split}
\int\phi^{(1)}\begin{pmatrix} 1\\ {\bf v}\\\frac{v^2}{2}\end{pmatrix}d^Dv&=-\tau\frac{\partial}{\partial t_1}\int\begin{pmatrix} 1\\ {\bf v}\\\frac{v^2}{2}\end{pmatrix}\phi^{(0)}d^Dv-\tau\nabla\cdot\int\begin{pmatrix} {\bf v}\\ {\bf vv}\\\frac{{\bf v}v^2}{2}\end{pmatrix}\phi^{(0)}d^Dv\\
\end{split}
\end{equation}
We define the mass density $\rho$, momentum density $\rho{\bf u}$, internal energy $\rho{\cal E}$:
\begin{equation}
\begin{split}
\rho&= \int \phi^0d^3v\\
\rho{\bf u} &= \int {\bf v}\phi^0d^3v\\
\rho{\cal E} &= \frac{1}{2}\int v^2\phi^0d^3v,\\
\end{split}
\end{equation}
and the inviscid pressure tensor ${\cal P}^{(0)}$ and heat flux $\rho{\bf J}^{(0)}$:
\begin{equation}
\begin{split}
{\cal P}^{(0)}&=\int {\bf vv}\phi^{(0)}d^3v\\
\rho{\bf J}^{(0)}&=\int \frac{v^2}{2}\phi^{(0)}{\bf v}d^3v\\
\end{split}
\end{equation}
and hence obtain the continuity and Eulers equations for the fluid.
\begin{equation}
\begin{split}
\frac{\partial\rho}{\partial t_1} + \nabla \cdot(\rho{\bf u})=0\\
\frac{\partial \rho{\bf u}}{\partial t_1}  + \nabla \cdot{\cal P}^0=0\\
\frac{\partial {\cal E}}{\partial t_1}  + \nabla \cdot\bigl({\rho {\bf J}^{(0)}}\bigr)=0\\
\end{split}
\end{equation}
\subsection{Second order solution}
We first simplify the equation for the second order deviation from the equilibrium using the first order equation:
\begin{equation}
\begin{split}
\phi^{(2)}&=-\tau({\cal D}_2-\tau{\cal D}_1^2)\phi^{(0)}\\
&=-\tau{\cal D}_2\phi^{(0)}-\tau{\cal D}_1\phi^{(1)}\\
\end{split}
\end{equation} 
Multiplying by the moments and integrating over velocities yields:
\begin{equation}
\begin{split}
\int\phi^{(2)}\begin{pmatrix} 1\\ {\bf v}\\\frac{v^2}{2}\end{pmatrix}d^Dv&=-\tau\frac{\partial}{\partial t_2}\int\begin{pmatrix} 1\\ {\bf v}\\\frac{v^2}{2}\end{pmatrix}\phi^{(0)}d^Dv-\tau\nabla\cdot\int\begin{pmatrix} {\bf v}\\ {\bf vv}\\\frac{{\bf v}v^2}{2}\end{pmatrix}\phi^{(0)}d^Dv\\
&-\tau\frac{\partial}{\partial t_1}\int\begin{pmatrix} 1\\ {\bf v}\\\frac{v^2}{2}\end{pmatrix}\phi^{(1)}d^Dv-\tau\nabla\cdot\int\begin{pmatrix} {\bf v}\\ {\bf vv}\\\frac{{\bf v}v^2}{2}\end{pmatrix}\phi^{(1)}d^Dv\\
\end{split}
\end{equation}
If we define the first order corrections to the pressure tensor and heat flux due to deviations from equilibrium:
\begin{equation}
\begin{split}
{\cal P}^{(1)}&=\int {\bf vv}\phi^{(1)}d^3v\\
\rho{\bf J}^{(1)}&=\int \frac{v^2}{2}\phi^{(1)}{\bf v}d^3v\\
\end{split}
\end{equation}
We obtain the viscous Navier-Stokes equations:
\begin{equation}
\begin{split}
\frac{\partial\rho}{\partial t_2} + \nabla \cdot(\rho{\bf u})=0\\
\frac{\partial \rho{\bf u}}{\partial t_2}  + \nabla \cdot\left({\cal P}^{(0)}+{\cal P}^{(1)}\right)=0\\
\frac{\partial \rho{\cal E}}{\partial t_2}  + \nabla \cdot\bigl(\rho {\bf J}^{(0)}+\rho {\bf J}^{(1)}\bigr)=0\\
\end{split}
\end{equation}
The expressions for the corrections to the pressure tensor and heat flux may be expressed in terms of gradients of the local equilibria by using the first order solution to our Boltzmann equation. Of course, the above equations will only be a useful fluid dynamical scheme if the pressure tensor and heat flux may be expressed entirely in terms of other macroscopic quantities. It is to this issue that we turn in the remainder of the paper.

\section{Moments of the distribution function}\label{sec:mom}

In this section we evaluate all moments of the distribution function defined above. The Chapman-Enskog expansion enables us to express deviations of the distribution function from equilibrium in terms of gradients of the local equilibrium distribution. All moments may therefore be evaluated in terms of the function $\phi^{(0)}$. As we shall see, the conditions of isotropy and Gallilean invariance, which this function must satisfy, are sufficient to enable us to evaluate all moments of this function in terms of the lowest order moments. These lowest order moments are precisely the  hydrodynamic variables, and so we obtain an autonomous macroscopic description under very mild assumptions about the form of the local equilibrium distribution.

Our results will follow from a remarkable property of the Taylor expansion of isotropic functions, which is proved in~\cite{bib:lovepart}. We first introduce some notation. We denote the $mth$ rank outer product of a vector ${\bf v}$ with itself by: 
\begin{equation}\nonumber
\begin{split}
\bigotimes^m{\bf v} &= \underbrace{{\bf vvvv}\cdots{\bf vvv}}\\
&\phantom{=vvvv}m~times
\end{split}
\end{equation}
similarly we denote the $mth$ rank outer product of gradient operators by:
\begin{equation}\nonumber
\begin{split}
\bigotimes^m{\bf \nabla} &= \underbrace{\nabla\nabla\nabla\cdots \nabla\nabla}\\
&\phantom{=\nabla\nabla}m~times
\end{split}
\end{equation}
this operator is equal to the familiar Hessian operator for $m=2$ and is the higher rank generalization of the Hessian for $m>2$.

We combine this notation for the outer product of vectors with a similar notation for the inner product (or more generally a p-fold contraction operation): 
\begin{equation}\bigotimes^r{\bf a}\bigodot^p\bigotimes^q{\bf b}=({\bf a}\cdot{\bf b})^p\bigotimes^{r-p}{\bf a}\bigotimes^{q-p}{\bf b}\end{equation}
The above signifies the $p$-fold contraction of a $qth$ rank tensor (in fact a $q$-adic) on a $rth$ rank tensor (an $r$-adic).  

We now write the usual Taylor expansion of a function of a vector:
\begin{equation}f({\bf x+a})=\sum_{n=0}^{{\cal 1}}({\bf a\cdot \nabla})^nf({\bf x})\end{equation}
We write this in the notation defined above as:
\begin{equation}f({\bf x+a})=\sum_{n=0}^{{\cal 1}}\bigotimes^n{\bf a}\bigodot^n\biggl[ \bigotimes^n\nabla f({\bf x})\biggr]\end{equation}

While it may appear that the purpose of this notation is to take the simple, well known expression for the Taylor expansion and convert it to a complicated and unwieldy one, we are not being deliberately obscurantist. The point of this notation is to associate with each order of the Taylor expansion the tensor:
\begin{equation}T_n=\biggl[ \bigotimes^n\nabla f({\bf x})\biggr].\end{equation}
We are interested in isotropic functions of vectors, that is, functions only of the modulus of the vector. As we shall see in a moment, the tensors $T_n$ associated with the Taylor expansion of such isotropic functions take a particularly simple form.

We follow the notation of~\cite{bib:lovepart} and define the $n$th-rank completely symmetric tensor kernel
\begin{equation}
\calk_n ({\bf v})
 \equiv
 \sum_{m=\lceil n/2 \rceil}^n
 \frac{\phi_m(v)}{(n-m)!}\,
 \mbox{\rm per}
 \left[
  \left(\bigotimes^{2m-n}{\bf v}\right)
  \otimes
  \left(\bigotimes^{n-m}\bfone\right)
 \right],
\label{eq:kernel}
\end{equation}
where ``per'' indicates a summation over all distinct permutations of
indices, and where we have defined the following functions related to
the derivatives of $\phi (v)$,
\begin{equation}
\phi_m(v)\equiv
 \left(\frac{1}{v}\frac{d}{dv}\right)^m
 \phi(v).
\end{equation}
 and state the following property:
\begin{equation}\label{eq:recurr}
\bigotimes^m \nabla\calk_n ({\bf v})=\calk_{n+m} ({\bf v})
\end{equation}
The first few such kernels are given in appendix~\ref{sec:kernels}

We may write the Mach number expansion of our distribution as a sum of complete contractions of totally symmetric $nth$ rank tensors:
\begin{equation}
\phi(|{\bf v}-{\bf u}|) = \sum_{n=0}^{\cal 1}\bigl(-1\bigr)^n\frac{1}{n!}\biggl[\bigotimes^n\nabla\phi(|{\bf v}|)\biggr]\bigodot^n\bigotimes^n{\bf u}
\end{equation}
Using our tensor kernels defined above and the identity~(\ref{eq:recurr}) we identify $\calk_0({\bf v})=\phi(|{\bf v}|)$ and write
\begin{equation}
\phi(|{\bf v}-{\bf u}|) = \sum_{n=0}^{\cal 1}\bigl(-1\bigr)^n\frac{1}{n!}\calk_n({\bf v})\bigodot^n\bigotimes^n{\bf u}
\end{equation}
For an isotropic function the tensors $T_n$ appearing in the Taylor expansion are exactly the completely isotropic tensors given by~(\ref{eq:kernel}). If one compares this expression for the Taylor expansion of a function of the modulus of a vector with that given in~\cite{bib:micrork} the economy of this notation becomes apparent.

We now define the moments of the distribution via the two sets of integrals:
\begin{equation}
I^m_n=\int_V {\cal K}_n({\bf v})\bigotimes^m{\bf v} d^Dv.
\end{equation}
and
\begin{equation}
J^m_n = \frac{1}{2}\int {\cal K}_n({\bf v})v^2\bigotimes^m{\bf v} d^Dv 
\end{equation}

These integrals have the property $I^m_n=0$ if $n>m$ and $J^m_n=0$ if $n>m+2$. Furthermore both integrals vanish if $m+n$ is odd. This implies that for each moment of the distribution function only a finite number of terms in the Mach number expansion is necessary to exactly compute that moment. Furthermore, as a consequence of~(\ref{eq:recurr}) the integrals above satisfy a relatively simple recursion relation which means that they may all be computed in terms of the lowest order integrals. 

We first consider the integrals $I^m_n$. First note that the integrand is odd for $m+n$ odd, and so $I^m_n$ vanishes for $m+n$ odd. The asymptotic behaviour of our tensors is governed by the set of functions $\phi_n$. The kernel ${\cal K}_n({\bf v})$ behaves asymptotically like $v^n\phi_n$. We require ${\cal K}_0({\bf v})$ to be normalisable, that is the integral:
\begin{equation}
\int {\cal K}_0({\bf v}) d^Dv,
\end{equation}
must converge. This means that if ${\cal K}_0({\bf v})$ is algebraic it must behave asymptotically like $\sim v^{-p}$ where $p > D-1$. The functions $\phi_n$ behave asymptotically like $v^{-(p+2n)}$, and hence the kernel ${\cal K}_n({\bf v})$ behaves asymptotically like $v^{-q}$ where $q=(p+n)$ and $q> n+D-1$. This implies that the integrals:
\begin{equation}
\int {\cal K}_n({\bf v})\bigotimes^m{\bf v}d^Dv,
\end{equation}
converge if $m<q$, i.e. if $m\leq n+D-1$. The first two integrals are determined by the fact that ${\cal K}_0$ is the isotropic part of the comoving distribution function:
\begin{equation}
\begin{split}
I^0_0 &=\rho\\
I^1_0 &=0
\end{split}
\end{equation}
We define the temperature as proportional to the average of the particles kinetic energy, where the proportionality is chosen in accordance with the equipartition theorem. 
\begin{equation}
\begin{split}
\frac{\rho DkT}{2} &= <\frac{1}{2}mv^2>=\frac{1}{2}\int \calk_0({\bf v})v^{D+1}dv\int d^D\Omega=\frac{S_D}{2}\int \calk_0({\bf v})v^{D+1}dv\\
\end{split}
\end{equation}Where $S_D$ is the surface area of the unit sphere in $D$ dimensions. This definition of temperature is microscopic, and hence valid out of equilibrium. This will correspond with the macroscopic equilibrium thermodynamic temperature if and only if the comoving distribution is Maxwellian. We impose no such restriction. 
Introducing the unit vector ${\bf e}_v$ in the ${\bf v}$ direction we write our first nontrivial tensor moment in terms of the temperature:
\begin{equation}
\begin{split}
I^2_0&=\int \calk_0({\bf v})v^{D+1}dv\int{\bf e}_v{\bf e}_vd^D\Omega=\frac{S_D}{D}{\bf 1}\int \calk_0({\bf v})v^{D+1}dv\\
\end{split}
\end{equation}
Where we have used:
\begin{equation}
\begin{split}
\int{\bf e}_v{\bf e}_vd^D\Omega&=\frac{S_D}{D}\bfone
\end{split}
\end{equation}
Giving:
\begin{equation}
I^2_0 = \rho kT {\bf 1} 
\end{equation}
We can now proceed to the general case. We state the following theorem:
\begin{equation}
\nabla \bigotimes^m {\bf v} = 
 \mbox{\rm per}
 \left[
  \left(\bigotimes^{m-1}{\bf v}\right)
  \otimes
  \left(\bfone\right)
 \right] - {\bf v} \mbox{\rm per}
 \left[
  \left(\bigotimes^{m-2}{\bf v}\right)
  \otimes
  \left(\bfone\right)
 \right]
\end{equation}
where by definition any outer tensor product of negative rank is zero. The proof of this theorem is given in~\cite{bib:lovepart}. We utilise this below:
\begin{equation}
\begin{split}
I^m_n &=\int_{V}\biggl\{\nabla\biggl[{\cal K}_{n-1}({\bf v})\bigotimes^m{\bf v}\biggr]-\biggl(\nabla\bigotimes^m{\bf v}\biggr){\cal K}_{n-1}({\bf v})\biggr\}d^Dv\\
& = {\cal K}_{n-1}({\bf v})\bigotimes^m{\bf v}\bigg|_{\partial V}-\int  {\cal K}_{n-1}({\bf v}) \left\{{\rm per}
 \left[
  \left(\bigotimes^{m-1}{\bf v}\right)
  \otimes
  \bfone
 \right] - {\bf v} \mbox{\rm per}
 \left[
  \left(\bigotimes^{m-2}{\bf v}\right)
  \otimes
  \bfone
 \right]
d^Dv\right\}\\
\end{split}
\end{equation}
The boundary term vanishes for $m \leq D+n-1$ and we obtain the 
slightly awkward recursion relation:
\begin{equation}
\begin{split}
I^m_n & = -\int  {\cal K}_{n-1}({\bf v}) \left\{{\rm per}
 \left[
  \left(\bigotimes^{m-1}{\bf v}\right)
  \otimes
  \bfone
 \right] - {\bf v} \mbox{\rm per}
 \left[
  \left(\bigotimes^{m-2}{\bf v}\right)
  \otimes
  \bfone
 \right]
d^Dv\right\}\\
\end{split}
\end{equation}
We write the recursion relation in terms of tensors $\widehat{per}[m,1]$, which are defined and discussed in detail in appendix~\ref{sec:appper}:
\begin{equation}
\begin{split}
I^m_n & = -\int  {\cal K}_{n-1}({\bf v}) \widehat{\rm per}[m-1,1]d^Dv\\
\end{split}
\end{equation}
We immediately obtain two results of importance. Firstly we note that $I^0_n=0$ for $n>0$. This property guarantees that the higher orders in the Mach number expansion of our equilibrium distribution do not contribute to the value of the density. Secondly, using the recursion relation it is clear that $I^0_n=0$ for $n>0$ implies the general result that $I^m_n=0$ for $n>m$.

We now consider the integrals $J^m_n$, which, by an extension of the argument given above for $I^m_n$, are convergent for $m\leq n+D -3$. Write:
\begin{equation}
\nabla\biggl[{\cal K}_{n-1}({\bf v})v^2\bigotimes^m{\bf v} \biggr] ={\cal K}_{n-1}({\bf v}) \nabla (v^2)\bigotimes^m{\bf v} +{\cal K}_{n}({\bf v})v^2\bigotimes^m{\bf v}+{\cal K}_{n-1}({\bf v})v^2\nabla\bigotimes^m{\bf v}
\end{equation}
giving:
\begin{equation}
J^m_n = \frac{1}{2}v^2{\cal K}_{n-1}({\bf v})\bigg|_{\partial V} - \int {\cal K}_{n-1}({\bf v})\bigotimes^{m+1}{\bf v} d^Dv - \frac{1}{2}\int (v^2){\cal K}_{n-1}({\bf v})\nabla\biggl(\bigotimes^{m}{\bf v} \biggr)d^Dv
\end{equation}
Introducing our tensors $\widehat{per}[m,1]$ we obtain:
\begin{equation}
J^m_n = - I^{m+1}_{n-1} - \frac{1}{2}\int (v^2){\cal K}_{n-1}({\bf v})\widehat{per}[m-1,1]d^Dv
\end{equation}
The temperature is defined via $J^0_0=D\rho kT/2$. The recursion above implies $J^0_n=0$ for $n>2$, and hence $J^m_n=0$ for $n> m+2$. 
Evaluation of the first integrals gives:
\begin{equation}
\begin{split}
J^0_2 &= -I^1_1 =\rho{\bf 1}\\
J^1_1 &=-I^2_0-{\bf 1}J^0_0=-\rho kT{\bf 1}-\frac{D\rho kT}{2}{\bf 1}= -\frac{\rho kT(D+2)}{2}{\bf 1}\\
[J^1_3]_{ijkl} &=-I^2_2 - {\bf 1}J^0_2=-\rho(\delta_{ik}\delta_{jl}+\delta_{il}\delta_{jk})-\rho\delta_{ij}\delta_{kl}=-\rho\Omega_{ijkl}\\
\end{split}
\end{equation}
Where $\Omega_{ijkl}$ is the completely isotropic fourth rank tensor defined in appendix~\ref{sec:intdetails}. We also require the integrals $J^2_0, J^2_2, J^2_4$. The integral $J^2_0$ does not follow from any of the integrals considered so far. This is because it is proportional to the fourth moment of the distribution function. For the Maxwell-Boltzmann distribution the fourth moment may be expressed in terms of the second moments of the distribution. For the quite general distributions considered here this may no longer be true, and we introduce a new scalar variable $R$:
\begin{equation}
R=\frac{1}{D(D+2)}\int v^4\phi d^Dv=\frac{1}{D(D+2)}\int v^{D+3}\phi dv\int d\Omega=\frac{S_D}{D(D+2)}\int v^{D+3}\phi dv
\end{equation} 
Where we have chosen the definition so that the related integral $I^4_0$ is independent of dimension. We may now write $J^2_0$ in terms of this new scalar variable:
\begin{equation}
\begin{split}
J^2_0&=\frac{1}{2}\int v^2{\bf vv} \phi d^Dv = \frac{1}{2}\int v^4\phi d^Dv=\frac{1}{2}\int v^{D+3}\phi dv\int {\bf e_ve_v}d\Omega\\
&=\frac{\bf 1}{2D}S_D\int v^{D+3}\phi dv=\frac{(D+2)}{2}R{\bf 1}\\
\end{split}
\end{equation}

All integrals may now be computed in terms of $\rho$, $T$, $R$ and ${\bf u}$. The remaining higher order integrals are used below and computed in detail in appendix~\ref{sec:intdetails}. We begin with the inviscid pressure tensor $\mathcal{P}^0$:
\begin{equation}
\begin{split}
{\cal P}^0&=\int \bigl({\bf vv}\phi^0_0\bigr)d^3v=I^2_0+\frac{\bf uu}{2}:I^2_2 =\rho kT{\bf 1} +\rho{\bf uu}\\
\end{split}
\end{equation}
the energy density and heat flux:
\begin{equation}
\begin{split}
\rho{\cal E} &= \frac{1}{2}\int v^2\phi^0d^3v= J^0_0+\frac{\bf uu}{2}:J^0_2\\
&=\frac{\rho}{2}\bigl(DkT+u^2)\\
\rho{\bf J}&=\frac{1}{2}\int \bigl(v^2{\bf v}\phi^0\bigr)d^3v=-{\bf u}\cdot J^1_1-\frac{\bf uuu}{3!}\odot^3 J^1_3\\
&=\frac{\rho kT(D+2)}{2}{\bf u}+\frac{\rho u^2}{2}{\bf u}
\end{split}
\end{equation}
Note that the internal energy is defined as:
\begin{equation}
\begin{split}
\rho{ i} &=\frac{\rho DkT}{2}\\
\end{split}
\end{equation}
and hence the ideal gas equation of state for the scalar pressure $P$ is independent of the form of the distribution function when written in terms of the internal energy:
\begin{equation}
\begin{split}
P&= \frac{2}{D}\rho{ i}
\end{split}
\end{equation}
The first correction is to the pressure tensor:
\begin{equation}
{\cal P}^{(1)} = \int \phi^{(1)}{\bf vv}d^Dv
\end{equation}
Substituting from above and using the integrals computed in appendix~\ref{sec:intdetails}:
\begin{equation}
\begin{split}
{\cal P}^{(1)}&=-\tau\frac{\partial {\cal P}^{(0)}}{\partial t_1}+\tau\nabla\cdot\left[{\bf u}\cdot I^3_1  +\frac{\bf uuu}{3!}\odot^3 I^3_3\right]\\
&=-\tau\frac{\partial {\cal P}^{(0)}}{\partial t_1}-\tau\left[\nabla\cdot\left(\rho kT{\bf u}\right) +\nabla\left(\rho kT{\bf u}\right) +\left[\nabla\left(\rho kT {\bf u}\right)\right]^T+\nabla\cdot({\bf \rho uuu})\right]\\
\end{split}
\end{equation}
The first order correction to the heat flux is then:
\begin{equation}
\begin{split}
\rho{\bf J}^{(1)}&=\int \frac{v^2}{2}{\bf v}\phi^{(1)}d^Dv\\
&=-\tau\frac{\partial}{\partial t_1}\left[{\bf u}\cdot J^1_1+\frac{\bf uuu}{3!}\odot^3 J^1_3\right] -\tau \nabla\cdot\left[J^2_0 +\frac{\bf uu}{2}:J^2_2+\frac{\bf uuuu}{4!}:J^2_4\right]\\
&=-\tau\frac{\partial}{\partial t_1}\left[\rho{\bf J}^{(0)}\right] -\tau \nabla\cdot{\cal Q}\\
\end{split}
\end{equation}
Where we have introduced the heat flux tensor for brevity. This quantity may be written out explicitly as:
\begin{equation}
\begin{split}
{\cal Q}&=\left(\frac{(D+2)}{2}R+\frac{Pu^2}{2}\right){\bf 1} +\left(\rho{\cal E} +2P\right){\bf uu}
\end{split}
\end{equation}
The only step which remains is to express the time derivatives with respect to $t_1$ in terms of spatial derivatives using the first order equations of motion. After some laborious but straightforward manipulations we obtain:
\begin{equation}
\begin{split}
{\cal P}^{(1)}&=-\left\{\mu[\nabla\left({\bf u}\right) +\left[\nabla\left( {\bf u}\right)\right]^T]+\lambda {\bf 1}\nabla\cdot{\bf u}\right\}\\
\end{split}
\end{equation}
Where we have introduced the shear viscosity:
\begin{equation}
\mu = \tau \rho kT 
\end{equation}
and the bulk viscosity
\begin{equation}
\lambda = -\tau \frac{2}{D}\rho kT 
\end{equation}
We also obtain the first order correction to the heat flux:
\begin{equation}
\begin{split}
\rho{\bf J}^{(1)}&=-{\bf u}\cdot\biggl(\mu[\nabla{\bf u}+(\nabla {\bf u})^T]+\lambda{\bf 1}\nabla\cdot{\bf u}\biggr)+ \tau\frac{(D+2)}{2}\biggl\{kT\nabla (\rho kT) -\nabla(\rho R) \biggr\}\\
\end{split}
\end{equation}
If we now define:
\begin{equation}\label{theta}
\theta = R-(kT)^2
\end{equation}
We may write:
\begin{equation}
\begin{split}
\rho{\bf J}^{(1)}&=-{\bf u}\cdot\biggl(\mu[\nabla{\bf u}+(\nabla {\bf u})^T]+\lambda{\bf 1}\nabla\cdot{\bf u}\biggr)- \tau\frac{(D+2)}{2}\biggl\{\rho kT\nabla (kT) +\nabla(\rho \theta) \biggr\}\\
&=-{\bf u}\cdot\biggl(\mu[\nabla{\bf u}+(\nabla {\bf u})^T]+\lambda{\bf 1}\nabla\cdot{\bf u}\biggr)- \kappa\nabla { i}-\bar\kappa\nabla(\rho \theta) \biggr\}\\
\end{split}
\end{equation}
Where we have defined the thermal conductivity:
\begin{equation}
\kappa = \frac{D+2}{2}\tau\rho {i}
\end{equation}
and the anomalous transport coefficient:
\begin{equation}
\bar\kappa = \frac{D+2}{2}\tau
\end{equation}
These transport coefficients have been defined such that the anomalous transport term vanishes when the quantity $\theta$, vanishes. This is the case if $\phi^{(0)}$ is a local Maxwellian, and if $\phi^{(0)}$ is a Tsallis distribution these results are identical to those obtained in~\cite{bib:bogtsallis}.

\section{Conclusions}

We have shown that the Navier-Stokes equations for incompressible flow are independent of the detailed form of the local distribution function, provided only that the distribution may be written as an isotropic function of velocity with small anisotropic corrections. The energy equation in a suitably generalized form is also independent, acquiring only one extra term for non-Maxwellian distributions. This extra term is the only additional term allowed by the Curie principle.

The original motivation of Boghosians derivation of hydrodynamic equations for systems with a local Tsallis equilibria was the potential for determining a systems thermostatistics by performing entirely hydrodynamic experiments. The present work extends this hope to the full range of non-Maxwellian distributions currently being considered, with the caveat that two distributions whose fourth velocity moments have the same dependendence on their lower order moments will be empirically equivalent in this regard.

We expect that the techniques employed in this paper will be of further assistance in the derivation and elucidation of more complex hydrodynamic schemes by the Chapman-Enskog method. In particular, obtaining such schemes for interacting fluids is fraught with difficulty, and usually requires one to make some {\em ansatz} for the equilibrium distribution of the fluid~\cite{bib:saurobook,bib:sauroref,bib:basleb2,bib:luobinary}. One might hope that by using methods which do not require specification of the detailed form of the local equilibrium, one might place such interacting fluid models on a firmer theoretical foundation.

\section*{Acknowledgements}

PJL and BMB would like to thank Sauro Succi and Li-Shi Luo for useful discussions. PJL was supported by the DARPA QuIST program under AFOSR grant number [F49620-01-1-0566]. BMB was supported in part by the U.S. Air Force Office of Scientific Research under grant number [F49620-01-1-0385]. 

%\bibliographystyle{prsty}
%\bibliography{my}

\appendix

\section{Completely symmetric tensor kernels}\label{sec:kernels}

The first few tensor kernels as defined in eq.~(\ref{eq:kernel}) are:
\begin{eqnarray}\nonumber
\left[\calk_1 (\bfy)\right]_{i}
 &=&
 \phi_1 (y) y_i
\\
\left[\calk_2 (\bfy)\right]_{ij}\nonumber
 &=&
 \phi_1 (y) \delta_{ij} +
 \phi_2 (y) y_iy_j
\\
\left[\calk_3 (\bfy)\right]_{ijk}\nonumber
 &=&
 \phi_2 (y) \left( y_i\delta_{jk} +
                   y_j\delta_{ik} +
                   y_k\delta_{ij}
            \right) +
 \phi_3 (y) y_iy_jy_k
\\
\left[\calk_4 (\bfy)\right]_{ijkl}
 \nonumber&=&
 \phi_2 (y) \left( \delta_{ij}\delta_{kl} +
                   \delta_{ik}\delta_{jl} +
                   \delta_{il}\delta_{jk}
            \right)/2 +
\nonumber\\
& &
 \phi_3 (y) \left( y_iy_j\delta_{kl} +
                   y_iy_k\delta_{jl} +
                   y_iy_l\delta_{jk} +
                   y_jy_k\delta_{il} +
                   y_jy_l\delta_{ik} +
                   y_ky_l\delta_{ij}
            \right) +
\nonumber\\
& &
 \phi_4 (y) y_iy_jy_ky_l.\nonumber
\end{eqnarray}

\section{Moments of flux field expansions}\label{sec:intdetails}

We first give the value of the following angular integral, which will be useful for the higher order integrals to follow:
\begin{equation}
\begin{split}
\int e_ie_je_ke_ld^D\Omega&= \frac{S_D}{D(D+2)}( \delta_{ij}\delta_{kl} +\delta_{ik}\delta_{jl} + \delta_{il}\delta_{jk})\\
&=\frac{S_D}{D(D+2)}\Omega_{ijkl}
\end{split}
\end{equation}
Where we have defined the completely isotropic fourth rank tensor $\Omega_{ijkl}$. We consider the integrals $I^0_1, I^1_1, I^2_1, I^3_1$. The integrals $I^0_1, I^2_1$ vanish by symmetry: 
\begin{equation}
\begin{split}
I^1_1&=-\int  {\cal K}_{0}({\bf v}) \widehat{\rm per}[0,1]d^Dv=-\rho{\bf 1}\nonumber
\end{split}
\end{equation}
and
\begin{equation}
\begin{split}\nonumber
I^3_1&=-\int  {\cal K}_{0}({\bf v}) \widehat{\rm per}[2,1]d^Dv\\
&=-\int \phi v^{D+1}dv\biggl(\int e_ke_ld^D\Omega \delta_{ij}+\int e_je_ld^D\Omega \delta_{ik}+\int e_je_kd^D\Omega \delta_{il}\biggr)\\
&=-\rho kT\Omega_{ijkl}
\end{split}
\end{equation}
We now consider the integrals $I^0_2, I^1_2, I^2_2, I^3_2, I^4_2$. The integrals  $I^1_2, I^3_2$ vanish by symmetry, and $I^0_n= 0$ for $n>0$, 
\begin{equation}
\begin{split}\nonumber
\biggl[I^2_2\biggr]_{hijk} &=-\int{\cal K}_1({\bf v})\widehat{per}[1,1]d^Dv\\
&=-\int \phi_1(v) v_h(v_j\delta_{ik} +v_k\delta_{ij})d^Dv\\
&=-\int \frac{1}{v}\frac{d\phi}{dv}v^2v^{D-1}dv\biggl( \delta_{ik}\int e_he_jd^D\Omega + \delta_{ij}\int e_he_kd^D\Omega\biggr)\\
&=S_D\int \phi v^{D-1}dv\biggl( \delta_{ik}\delta_{hj} + \delta_{ij}\delta_{hk}\biggr)\\
&=\rho\biggl( \delta_{ik}\delta_{hj} + \delta_{ij}\delta_{hk}\biggr)\\
\end{split}
\end{equation}
and
\begin{equation}
\begin{split}\nonumber
\biggl[I^4_2\biggr]_{efhijk} &=-\int{\cal K}_1({\bf v})\widehat{per}[3,1]d^Dv\\
&=-\int \phi_1(v) v_e(v_iv_jv_k\delta_{hl} +
                   v_iv_kv_l\delta_{jh} +
                   v_iv_lv_j\delta_{hk} +
                   v_jv_kv_l\delta_{ih})d^Dv\\
&=-\int \frac{1}{v}\frac{d\phi}{dv}v^4v^{D-1}dv\biggl( \delta_{hl}\int e_ee_ie_je_kd^D\Omega + \delta_{jh}\int e_ee_ie_ke_ld^D\Omega\\
&+\delta_{hk}\int e_ee_ie_le_jd^D\Omega + \delta_{ih}\int e_ee_je_ke_ld^D\Omega\biggr)\\
\end{split}
\end{equation}
We know the result of the integration over solid angle and integrate over the speed by parts:
\begin{equation}
\begin{split}\nonumber
\biggl[I^4_2\biggr]_{efhijk} &=\frac{S_D}{D}\int \phi v^{D+1}dv\biggl( \delta_{hl}\Omega_{eijk} + \delta_{jh}\Omega_{eikl}+\delta_{hk}\Omega_{eilj} + \delta_{ih}\Omega_{ejkl}\biggr)\\
\end{split}
\end{equation}
Substituting again from our definition of temperature we obtain:
\begin{equation}
\begin{split}\nonumber
\biggl[I^4_2\biggr]_{efhijk} &=\rho kT\biggl( \delta_{hl}\Omega_{eijk} + \delta_{jh}\Omega_{eikl}+\delta_{hk}\Omega_{eilj} + \delta_{ih}\Omega_{ejkl}\biggr)\\
\end{split}
\end{equation}
Finally, we consider $I^3_3$, the final integral required in the analysis above. 
\begin{equation}
\begin{split}\nonumber
I^3_3&=-\int {\cal K}_2({\bf v})\widehat{per}[2,1]d^Dv\\
\left[I^3_3\right]_{ijklmn}&=-\left[I^2_2\right]_{ijmn}\delta_{kl} -\left[I^2_2\right]_{ijln}\delta_{km}-\left[I^2_2\right]_{ijlm}\delta_{kn}\\
&=-\rho\left[\delta_{jn}\delta_{im}\delta_{kl}+\delta_{jm}\delta_{in}\delta_{kl}+\delta_{jn}\delta_{il}\delta_{km}+\delta_{jl}\delta_{in}\delta_{km}+\delta_{jm}\delta_{il}\delta_{kn}+\delta_{jl}\delta_{im}\delta_{kn}\right]   
\end{split}
\end{equation}
We now turn our attention to the integrals $J^2_2, J^4_2$, which may be defined in terms of previously computed integrals:
\begin{equation}
\begin{split}\nonumber
J^2_2&=-I^3_1-\frac{1}{2}\int v^2 {\cal K}_1({\bf v})\widehat{per}[1,1]d^Dv\\
\left[J^2_2\right]_{ijkl}&=\rho kT\Omega_{ijkl}-\frac{1}{2}\left(J^1_1\right)_{ik}\delta_{jl} -\frac{1}{2}\left(J^1_1\right)_{il}\delta_{jk}\\
&=\rho kT\left[\Omega_{ijkl} +\frac{(D+2)}{4}\left(\delta_{ik}\delta_{jl}+\delta_{il}\delta_{jk}\right)\right]
\end{split} 
\end{equation}
and finally $J^2_4$:
\begin{equation}
\begin{split}\nonumber
J^2_4&=-I^3_3-\frac{1}{2}\int v^2 {\cal K}_3({\bf v})\widehat{per}[1,1]d^Dv\\
\left[J^2_4\right]_{ijklmn}&=-\left[I^3_3\right]_{ijklmn}-\left[J^1_3\right]_{ijkm}\delta_{ln} -\left[J^1_3\right]_{ijkn}\delta_{lm}\\
&=\rho\left[\delta_{jn}\delta_{im}\delta_{kl}+\delta_{jm}\delta_{in}\delta_{kl}+\delta_{jn}\delta_{il}\delta_{km}+\delta_{jl}\delta_{in}\delta_{km}+\delta_{jm}\delta_{il}\delta_{kn}+\delta_{jl}\delta_{im}\delta_{kn}\right]\\
&+\frac{\rho}{2}\left[\Omega_{ijkm}\delta_{ln} +\Omega_{ijkn}\delta_{lm}\right]
\end{split} 
\end{equation}

\section{Permutation operators}\label{sec:appper}
We have introduced the permutation operator $per$ via:
\begin{equation}\nonumber
 \mbox{\rm per}
 \left[
  \left(\bigotimes^{n}\bfy\right)
  \otimes
  \left(\bigotimes^{m}\bfone\right)
 \right]=\mbox{\rm per}
 \left[n,m \right]
\end{equation}
where ``per'' indicates a summation over all distinct permutations of
indices. The first few such tensors are thus given by
\begin{eqnarray}
\mbox{\rm per}\left[0,1 \right]\nonumber
 &=&
\delta_{ij}
\\
\mbox{\rm per}\left[0,2 \right]\nonumber
 &=&
 \delta_{ij}\delta_{kl} +
                   \delta_{ik}\delta_{jl} +
                   \delta_{il}\delta_{jk}
\\
\mbox{\rm per}\left[1,0 \right]\nonumber
&=&
 y_i
\\
\mbox{\rm per}\left[1,1 \right]\nonumber
 &=&
  y_i\delta_{jk} +
                   y_j\delta_{ik} +
                   y_k\delta_{ij}
\\
\mbox{\rm per}\left[2,0 \right]\nonumber
&=&y_iy_j
\\
\mbox{\rm per}\left[2,1 \right]\nonumber
&=&
y_iy_j\delta_{kl} +
                   y_iy_k\delta_{jl} +
                   y_iy_l\delta_{jk} +
                   y_jy_k\delta_{il} +
                   y_jy_l\delta_{ik} +
                   y_ky_l\delta_{ij}
\\ 
\mbox{\rm per}\left[3,0 \right]\nonumber
&=&
 y_iy_jy_k
\\
\mbox{\rm per}\left[4,0 \right]\nonumber
&=&
y_iy_jy_ky_l.
\end{eqnarray}
We only need to consider permutations of the form $per[n,1]$ for the evaluation of our recursion relation. These permutations are:
\begin{eqnarray}
\mbox{\rm per}\left[0,1 \right]\nonumber
 &=&
\delta_{ij}\nonumber
\\
\mbox{\rm per}\left[1,1 \right]\nonumber
 &=&
  y_i\delta_{jk} +
                   y_j\delta_{ik} +
                   y_k\delta_{ij}
\\
\mbox{\rm per}\left[2,1 \right]\nonumber
&=&
y_i\left(y_j\delta_{kl} +
                   y_k\delta_{jl} +
                   y_l\delta_{jk}\right) +
		y_ky_l\delta_{ij}+
		y_jy_l\delta_{ik} +
                   y_jy_k\delta_{il} 
\\
\mbox{\rm per}\left[3,1 \right]\nonumber
&=&
y_h\left(y_iy_j\delta_{kl} +
                   y_iy_k\delta_{jl} +
                   y_iy_l\delta_{jk} +
                   y_jy_k\delta_{il} +
                   y_jy_l\delta_{ik} +
                   y_ky_l\delta_{ij}\right)+\\\nonumber
& &
		\left(y_iy_jy_k\delta_{hl} +
                   y_iy_ky_l\delta_{jh} +
                   y_iy_ly_j\delta_{hk} +
                   y_jy_ky_l\delta_{ih}\right)
\\
\mbox{\rm per}\left[4,1 \right]\nonumber
&=&
y_f(y_hy_iy_j\delta_{kl} +
                   y_hy_iy_k\delta_{jl} +
                   y_hy_iy_l\delta_{jk} +
                   y_hy_jy_k\delta_{il} +
                   y_hy_jy_l\delta_{ik} +\\\nonumber
& &
                   y_hy_ky_l\delta_{ij}+
		y_iy_jy_k\delta_{hl} +
                   y_iy_ky_l\delta_{jh} +
                   y_iy_ly_j\delta_{hk} +
                   y_jy_ky_l\delta_{ih})+\\\nonumber
& &
                   y_iy_jy_ky_l\delta_{hf} +
                   y_hy_jy_ky_l\delta_{if} +
                   y_hy_iy_ky_l\delta_{jf} +
                   y_hy_iy_jy_l\delta_{kf} +
                   y_hy_iy_jy_k\delta_{lf}. 
\end{eqnarray}
We have positioned the brackets above to illustrate the pattern which occurs among these tensors. Let $P(n)$ be the number of terms in the permutation $per[n,1]$. In the permutation $per[n+1,1]$ there will be $P(n)$ terms in which the new index is attached to one of the ${\bf y}$'s. $per[n+1,1]$ is an $n+2th$ rank tensor and so there are $n+2$ distinct ways of attaching the new index to the kronecker delta. Therefore the number of terms in $per[n+1,1]$ is $P(n+1)=P(n) + n + 2$. 

Our second permutation operation is defined in terms of the above operation:
\begin{equation}\nonumber
\widehat{per}[n,1] = per[n,1] - {\bf y}per[n-1,1].
\end{equation}
Given the above decomposition of the $per$ operation, it is straightforward to write down the first few such tensors:
\begin{eqnarray}
\widehat{\rm per}\left[0,1 \right]_{ij}\nonumber
 &=&
\delta_{ij}\nonumber
\\
\widehat{\rm per}\left[1,1 \right]_{ijk}\nonumber
 &=&
                   y_j\delta_{ik} +
                   y_k\delta_{ij}
\\
\widehat{\rm per}\left[2,1 \right]_{ijkl}\nonumber
&=&
		y_ky_l\delta_{ij}+
		y_jy_l\delta_{ik} +
                   y_jy_k\delta_{il} 
\\
\widehat{\rm per}\left[3,1 \right]_{hijkl}\nonumber
&=&
		\left(y_iy_jy_k\delta_{hl} +
                   y_iy_ky_l\delta_{jh} +
                   y_iy_ly_j\delta_{hk} +
                   y_jy_ky_l\delta_{ih}\right)
\\
\widehat{\rm per}\left[4,1 \right]_{fhijk}\nonumber
&=&
                   y_iy_jy_ky_l\delta_{hf} +
                   y_hy_jy_ky_l\delta_{if} +
                   y_hy_iy_ky_l\delta_{jf} +
                   y_hy_iy_jy_l\delta_{kf} +
                   y_hy_iy_jy_k\delta_{lf}. 
\end{eqnarray}
Let $\widehat{P}(n)$ be the number of terms in $\widehat{per}[n,1]$. Using our previous recursion relation for $P(n)$ it is straightforward to see that $\widehat{P}(n)=n+1$. 

\end{document}